\documentclass[12pt]{article} 
\usepackage{a4,epsf,cite} 
 
\newcommand{\ve}[1]{\mbox{\boldmath$#1$}}

\title{Sedimentation of strongly and weakly charged colloidal particles:
       Prediction of fractional density dependence
      }

\vspace{0.5cm}

\author{Martin Watzlawek\thanks{
           corresponding author; 
           {\tt martin@thphy.uni-duesseldorf.de}
         }\\ 
        {\small Institut f{\"u}r Theoretische Physik II, 
                Heinrich-Heine Universit{\"a}t}
        \\ 
        {\small Universit{\"a}tsstr. 1, D-40225 D{\"u}sseldorf, Germany} 
        \\~\\
        and
        \\~\\
        Gerhard N{\"a}gele \\
        {\small Fakult{\"a}t f{\"u}r Physik, Universit{\"a}t Konstanz}
        \\
        {\small Postfach 5560, D-78434 Konstanz, Germany}
        }

\date{({\em Journal of Colloid and Interface Science} {\bf 214}, 170 (1999))}

\begin{document}    

\maketitle

\vfill
 
\begin{small}
   \noindent
   {\bf Key Words:} Sedimentation, Hydrodynamic Interactions, 
                    Charge-stabilized Colloids
\end{small}

\begin{small}
   \noindent
   {\bf Short running title:} Sedimentation of charged particles
\end{small}

\vfill

\pagebreak

\begin{abstract}
   We report on calculations of the reduced sedimentation velocity $U/U_{0}$
   in homogenous suspensions of strongly and weakly charged colloidal spheres
   as a function of particle volume fraction $\phi$. 
   For dilute suspensions of strongly charged spheres at low salinity,
   $U/U_{0}$ is well represented by the parametric form 
   $1 - p \, \phi^\alpha$ with a
   fractional exponent $\alpha=1/3$ and a parameter $p\simeq 1.8$, which is 
   essentially independent from the macroion charge $Z$.
   This non-linear
   volume fraction dependence can be {\em quantitatively} 
   understood in terms of a
   model of effective hard spheres with $\phi$-dependent diameter. 
   For weakly charged spheres in a deionized solvent, 
   we show that the exponent 
   $\alpha$ can be equal to $1/2$,
   if an expression for $U/U_0$ given by Petsev and Denkov 
   [{\em J. Colloid Interface Sci.} {\bf 149}, 329 (1992)] is employed. 
   We further show that the range
   of validity of this expression is limited to very small values of $\phi$ 
   and $Z$, which are probably not accessible in 
   sedimentation experiments. The presented results might also hold for
   other systems like spherical proteins or ionic micelles.
\end{abstract}

%
%

\section{Introduction}  
  
The sedimentation velocity $U$ of interacting colloidal particles depends 
both on the indirect hydrodynamic interactions (HI) 
mediated by the suspending solvent, 
and on the microstructure of the suspension. In equilibrium, 
the latter is determined by direct potential forces arising 
for example from the steric repulsion between the
particles and from the electrostatic repulsion of overlapping double layers.
Different pair potentials $u(r)$ lead to rather different 
microstructures, and this in turn strongly effects the sedimentation. 
It is well established by theories and experiments
that the sedimentation velocity of a dilute suspension of monodisperse
hard spheres is given by 
\begin{equation} \label{U.hs}
   \frac{U}{U_0}=1-6.55\phi+{\cal O}(\phi^2),
\end{equation}
where $\phi$ is the particle volume fraction, and 
$U_0$ is the sedimentation velocity at infinite dilution 
\cite{Batchelor:72,Cichocki:89,Clerx:92,Kops-Werkhoven:81,
      Buscall:82,ThiesWeesie:96,Xu:98}.
On the other hand, the long-ranged 
electrostatic repulsion occurring in 
suspensions of charged particles can give rise to a 
reduction in $U$, as compared 
to a hard sphere dispersion at the same volume fraction
\cite{Reed:Anderson:76,Reed:Anderson:80,Snook:84,Dickinson:80}. This decrease
in $U$ is mainly due to the cumulative backflow of displaced fluid, 
which becomes particularly effective because 
the probability of two or more charged particles coming close
to each other is very small. 
Conversely, hard spheres are effectively attracted to 
each other at small interparticle distances $r\stackrel{<}{\sim} 3a$,
where $a$ is the radius of the spheres. This 
can be readily seen from the potential of mean force $w(r)=-k_BT\ln g(r)$, 
where $g(r)$ is the radial distribution function of hard spheres, which
shows its maximum
value at contact distance $r=2a$ due to excluded volume effects. 
Closely spaced particles are mutually exposed to the downflow of nearby fluid,
dragged along with the sedimenting particles. Consequently, the retardation
from backflow is reduced, whereas the influence of near-field HI is enhanced
for hard sphere suspensions.

While the reduction in $U$ due to long-ranged electrostatic forces is 
known since 
many years, it was not realized until recently that, in particular for 
deionized charge-stabilized suspensions, 
$U$ can be significantly smaller than $U_0$ even for extremely small volume
fractions as $\phi\simeq10^{-4}$. In fact, in the past the
effects of HI have been frequently considered to be negligible for dilute
suspensions of charged particles \cite{Pusey:Tough:85,Pusey:91}. However, 
recent calculations have clearly demonstrated for these systems 
that HI is of importance at small $\phi$ not only for sedimentation
\cite{Naegele:94:1,Naegele:Habil:published,ThiesWeesie:95}, 
but also for short-time \cite{Naegele:93,Naegele:95:1} and
long-time \cite{Baur:96:1} collective diffusion, and for 
long-time self-diffusion \cite{Naegele:97:1}.
Moreover, it was shown theoretically for dilute suspensions of charged 
colloids without added electrolyte that $U$ follows a non-linear 
$\phi$-dependence of the parametric form
\cite{Naegele:94:1,Naegele:95:1,Naegele:Habil:published} 
\begin{equation} \label{U.charged}
   \frac{U}{U_0}=1-p\phi^{\frac{1}{3}}. 
\end{equation}
The numerically calculated coefficient $p\simeq1.8$ was found to be nearly 
independent of the (effective) particle charge $Z$, provided that $Z$ is kept 
large enough to completely mask the hard core of the particles.
Experimenal results of sedimentation experiments on charged colloids
agree favorable with the scaling-prediction of eq. (\ref{U.charged})
\cite{ThiesWeesie:95, Ackerson:private}. 
Similar non-linear volume fraction dependencies are found for the 
short-time translational and rotational diffusion coefficients
of charged particles \cite{Watzlawek:96:1,Watzlawek:97:1,Watzlawek:97:2}.

The same $\phi$-dependence as found for salt-free fluid suspensions 
of charged colloids is known to be valid for the sedimentation velocity
of dilute ordered arrays of fixed spheres 
\cite{Hasimoto:59,Saffman:73,Zick:Homsy:82,Brady:88}.
For such arrays, the coefficient $p$
is determined analytically as $p=1.76$ for a $sc$ lattice, and
as $p=1.79$ for a $fcc$ or a $bcc$ lattice \cite{Hasimoto:59, Zick:Homsy:82}.
As discussed in detail in Refs. 
\cite{Naegele:94:1,Naegele:95:1,Naegele:Habil:published}, 
the main peak position $r_{m}$ of the radial distribution function
$g(r)$ for highly charged particles in salt-free suspension scales
as $r_{m}\propto\phi^{-1/3}$ which is also typical of a crystalline solid.
This, in fact, turns out to be relevant for eq. (\ref{U.charged}) to be
valid for strongly repelling particles like charged colloids  
and for ordered arrays of fixed spheres
\cite{Naegele:94:1,Naegele:95:1,Naegele:Habil:published,Saffman:73}.
 
In the present article, we analyse the sedimentation velocity of
charged colloidal dispersions at low salinity both as a function of $\phi$
and of $Z$. Our numerical results for $U/U_{0}$ with system parameters
representing monodisperse modified PMMA particles 
investigated very recently in sedimentation
experiments \cite{Ackerson:private}, are well described by  
eq. (\ref{U.charged})
if the effective macroion charge is chosen such that $Z>100$.
We physically explain the scaling relation
eq. (\ref{U.charged}) in terms of an effective hard sphere (EHS) model by 
using Wertheims
analytical expression \cite{Wertheim:63} for the static structure 
function of hard
sphere dispersions obtained in Percus-Yevick approximation. 
In this specific form of the EHS model, also the numerically 
determined value $p\simeq 1.8$ is recovered very accurately.
We will further
demonstrate for weakly charged spheres that $U/U_0-1$ displays a volume
fraction dependence proportional to $\phi^{1/2}$, 
provided the suspension is completely deionized 
and $\phi$ is chosen small enough that the radial distribution function
of the macroions can be approximated by its zero-density limit.
This result, however, turns out to be valid only 
for extremely small values of $\phi$ and $Z$, which we believe 
are not accessible in sedimentation experiments.

\section{Theory of sedimentation in charged colloids}
\label{theory,section}

We start this section by summarizing the theoretical method used to 
calculate the reduced sedimentation velocity $U/U_0$ of charged colloidal spheres.
Our results for $U/U_0$ are based on the effective macroion fluid
model of charge-stabilized suspensions \cite{Naegele:Habil:published}. 
In this model, the effective
pair potential $u(r)$ between two charged colloidal particles
consists of a hard-core part with radius $a$, and of a longer-ranged screened
Coulomb potential $u_{el}(r)$ for $r>2a$, i.e.
\begin{equation} \label{yukawa.potential}
   \beta u_{el}(r)=2Ka\frac{e^{-\kappa(r-2a)}}{r}.
\end{equation}
Here, $K$ is a dimensionless coupling parameter given by
\begin{equation} \label{yukawa.coupling}
   K=\frac{L_B}{2a}\left(\frac{Z}{1+\kappa a}\right)^2,
\end{equation}
where $L_B=\beta e^2/\epsilon$ is the so-called 
Bjerrum length, and $\beta=(k_BT)^{-1}$
is the thermal energy.
The suspending fluid is treated as a continuum without internal structure, 
only characterized by its dielectric constant $\epsilon$. Moreover, $Z$ is the 
effective charge of a colloidal particle in units of the elementary charge
$e$. The screening parameter $\kappa$ is given by 
\begin{equation}
   \kappa^2=4\pi L_B\left[ n|Z|+2n_s\right]=\kappa_c^2+\kappa_s^2,
\end{equation}
where $n_s$ is the number density of added 1-1-electrolyte, and
$n=3\phi/(4\pi a^3)$ is the number density of colloidal particles.
Notice that $\kappa$ comprises a contribution 
$\kappa_c$
due to counterions, which are assumed to be monovalent, and a second 
contribution
$\kappa_s$ arising from added electrolyte.
Eq. (\ref{yukawa.potential}) is a good approximation for $u(r)$
even for strongly charged colloids and for values of $\kappa a$ significantly
larger than one, provided the effective charge number $Z$ is regarded as an 
adjustable
parameter \cite{Naegele:Habil:published}. 

On the time scales probed by dynamic light scattering and by sedimentation
experiments, the effect of HI on the translational motion of the colloidal
particles is embodied in the hydrodynamic diffusivity tensors 
$\ve{D}^{tt}_{ij}(\ve{R}^N)$; $i,j=1,\ldots,N$ 
\cite{Jones:Pusey:91,Naegele:Habil:published,Pusey:Tough:85,Pusey:91,Dhont:book}.
The short-hand notation $\ve{R}^N=(\ve{R}_1, \ldots,\ve{R}_N)$ denotes
the configuration of the $N$ spherical particles. 
Without HI,
$\ve{D}_{ij}^{tt}(\ve{R}^N)=\delta_{ij}D_0 {\bf 1}$, where 
$D_0=k_BT/(6\pi\eta a)$ is the Stokesian diffusion coefficient of 
a particle with radius $a$
in a solvent with viscosity $\eta$, and ${\bf 1}$
denotes the unit tensor. 

The reduced short-time sedimentation velocity
$U/U_0$ of a macroscopically homogeneous suspension of monodisperse colloidal 
spheres is then given by the zero-wavenumber limit 
\begin{equation} \label{U.lim.H}
   \frac{U}{U_0}=\lim_{q\to0}H(q)
\end{equation}
of the so-called hydrodynamic function $H(q)$ 
\cite{Naegele:Habil:published, Pusey:91, Naegele:93}. 
This function is defined as
\begin{equation} \label{Hq}
   H(q)=\frac{1}{ND_0} \sum^N_{l,j=1}  
        \left< 
           \hat{\ve{q}}\cdot\ve{D}^{tt}_{lj}(\ve{R}^N)\cdot\hat{\ve{q}}~ 
           e^{i\mbox{\scriptsize\boldmath$q$}\cdot
                (\mbox{\scriptsize\boldmath$R$}_l
                 -\mbox{\scriptsize\boldmath$R$}_j)} 
        \right>  
\end{equation}
with wavevector $\ve{q}$ of magnitude $q$ and corresponding unit vector 
$\hat{\ve{q}}=\ve{q}/q$. The brackets indicate an equilibrium ensemble average.
$H(q)$ can be regarded as a generalized (short-time) sedimentation
coefficient of particles exposed to spatially periodic external forces
aligned with $\hat{\ve{q}}$, and derived from a weak potential 
proportional to $\exp\left[-i\ve{q}\cdot\ve{r}\right]$ 
\cite{Russel:81,Naegele:97:1}.


In principle, one needs to distinguish the short-time sedimentation
velocity $U$, defined through eq. (\ref{U.lim.H}), from the long-time
sedimentation velocity, which is determined in conventional
sedimentation experiments. In dilute suspensions, however, when
the HI are well described as a sum of pairwise additive interactions, 
both quantities are identical,
since then sedimentation does not perturb the equilibrium microstructure
\cite{Russel:book,Glendinning:82,Dhont:book}. At larger
volume fractions, the microstructure becomes
distorted from its equilibrium form, since
$n$-body HI with $n\ge3$ becomes important. This distortion 
leads to an additional change in the sedimentation velocity.
Nevertheless, simulations of hard sphere suspensions show
that the differences between the short-time and the long-time
sedimentation velocities due to memory effects are rather small
\cite{Ladd:90,Ladd:93}. 

The assumption of pairwise additive HI is justified for the important
case of dilute (typically $\phi\le0.1$) charge-stabilized suspensions at
sufficiently low ionic strength, as considered here. For such systems, the
particles are kept apart from each other due to their strong electrostatic 
repulsion. Then the $N$-body diffusivity tensors $\ve{D}^{tt}_{ij}(\ve{R}^N)$
are well approximated by the two-body tensors
\begin{equation} \label{Dtt.pairwise}
   \ve{D}^{tt(2)}_{ij}(\ve{R}^N)=\delta_{ij}
      \left[
         {\bf 1} + \sum_{l=1}^{N}\,\!'\:\ve{A}(\ve{R}_i-\ve{R}_l)
      \right]+
      (1-\delta_{ij})\ve{B}(\ve{R}_i-\ve{R}_j),
\end{equation}
where the term $l=i$ is excluded from the sum.
This approximation has been verified in the case of translational and
rotational self-diffusion by considering also the leading three-body 
contribution to HI \cite{Watzlawek:96:1,Watzlawek:97:2},
and in case of the sedimentation velocity by comparison with a more elaborate
method, known as the lowest order form of the $\delta\gamma$-expansion
\cite{Beenakker:Mazur:84,Naegele:94:1,Genz:91}.
The two-body translational mobility
tensors $\ve{A}(\ve{r})$ and $\ve{B}(\ve{r})$ are calculated by means of
series expansions in powers of $(a/r)$ 
\cite{Jones:Schmitz:88,Cichocki:Felderhof:Schmitz:88}.
We only quote the leading terms for further reference
\begin{eqnarray}
   \label{series.A}
   \ve{A}(\ve{r})&=&
      -\frac{15}{4}\left(\frac{a}{r}\right)^4\hat{\ve{r}}\hat{\ve{r}}
      +{\cal O}(r^{-6})
\\ \label{series.B}
   \ve{B}(\ve{r})&=&
      \frac{3}{4}\left(\frac{a}{r}\right)
         \left[\ve{1}+\hat{\ve{r}}\hat{\ve{r}}\right]
      +\frac{1}{2}\left(\frac{a}{r}\right)^3
         \left[\ve{1}-3\hat{\ve{r}}\hat{\ve{r}}\right]
      +{\cal O}(r^{-7}),
\end{eqnarray}
with $\hat{\ve{r}}=\ve{r}/r$.
Using eqs. (\ref{Hq},\ref{Dtt.pairwise}), $H(q)$ is expressed in 
terms of an integral
\begin{equation} \label{H.pairlevel}
   H(q)=1+n\int d\ve{r}~g(r)
           \left[
              \hat{\ve{q}}\cdot\ve{A}(\ve{r})\cdot\hat{\ve{q}}+
              \hat{\ve{q}}\cdot\ve{B}(\ve{r})\cdot\hat{\ve{q}}
              \cos (\ve{q}\cdot\ve{r})
           \right],
\end{equation}
involving these mobility tensors together with the radial distribution
function $g(r)$. In this work, 
we use eqs. (\ref{U.lim.H},\ref{H.pairlevel})
together with the series expansions of $\ve{A}(\ve{r})$ and 
$\ve{B}(\ve{r})$ for calculating $U/U_0$, by including contributions
up to order $(a/r)^{20}$.

When terms only up to ${\cal O}(r^{-4})$ in the series expansions
of $\ve{A}(\ve{r})$ and $\ve{B}(\ve{r})$ are employed, $U/U_0$ is given
explicitly as
\begin{equation} \label{U.linear.phi}
   \frac{U}{U_0}=1-\phi\left[
                   5+3\int_2^\infty dx~x(1-g(x))
                   +\frac{15}{4}\int_2^\infty dx~\frac{g(x)}{x^2}
                \right],
\end{equation}
with $x=r/a$.

For charge-stabilized suspensions, it is not necessary to account for many 
terms in the expansion of the two-body mobility tensors, since the integrals
in eq. (\ref{H.pairlevel}) converge rapidly because $g(r)$ is essentially
zero at small interparticle distances. 
In contrast, many terms are needed for hard spheres.
For example, using the zero-density form $g_0(r)=\Theta(r-2a)$ for the
radial distribution function of hard spheres, we obtain from eq.
(\ref{U.linear.phi}) that $U/U_0=1-p\phi$ with $p=6.87$. Here,
$\Theta(x)$ is the
unit step function. By including terms only 
up to ${\cal O}(r^{-3})$, the result is $p=5.0$, as can be seen 
from eq. (\ref{U.linear.phi}), when the last term on the right hand side, 
arising from the term proportional to $(a/r)^4$
in eq. (\ref{series.A}) is omitted. On the other hand, if terms up to  
${\cal O}(r^{-20})$ are considered, the result for $p$ is 
improved to $p=6.54$, which
is close to the exact value $p=6.55$,
first obtained by Batchelor using tabulated numerical 
results for the near-field HI \cite{Batchelor:72}.

To obtain $U/U_0$ for dilute charge-stabilized suspensions, we calculate
$g(r)$ in the effective macroion fluid model by using, for simplicity,
the well-established rescaled mean spherical approximation
(RMSA)\cite{Naegele:Habil:published}. On the basis of the pairwise additivity
approximation of the HI combined with the $(a/r)$-expansion of the 
mobility tensors, henceforth 
referred to as PA-scheme, $U/U_0$ can then be calculated
as a function of $\phi$. 

\section{Fractional density dependence of $U/U_{0}$}
\label{fractional.section}

\subsection{Strongly charged particles}

We show in the following that the exponent $1/3$ in 
eq. (\ref{U.charged}) and the
charge-independence of the parameter $p$, both found from 
numerical calculations, can be understood {\em quantitatively} 
in terms of a model of effective hard spheres (EHS), which can be treated
{\em analytically}.

The EHS model accounts for the most important feature of the radial
distribution function $g(r)$ of highly charged particles, namely the
so-called correlation hole. For an illustration of that, consider
fig. \ref{gr.rmsa.ehs.plot}, which shows a typical $g(r)$
for a salt-free suspension of strongly charged particles at $\phi=0.08$.
%
%
%
%
Due to the strong electrostatic repulsion between the particles, 
$g(r)$ has a well developed first maximum, and a spherical region with 
nearly zero probability
of finding another particle. This region is referred to as
the correlation hole. 
For small $\phi$, the correlation hole usually 
extends over several particle diameters 
\cite{Naegele:Habil:published, Watzlawek:96:1}. Therefore, we can approximate 
the actual $g(r)$ of the charge-stabilized system by the
radial distribution function
of an effective hard sphere (EHS) system with
an effective
radius $a_{EHS}>a$ and an effective volume fraction
$\phi_{EHS}=\phi{(a_{EHS}/a)}^{3}$. 
The EHS-radius $a_{EHS}$ accounts for
the electrostatic repulsion between the particles and can be identified as
$a_{EHS}=r_{m}/2$, where $r_{m}$ is the principal peak position of the actual
$g(r)$. 

Since the extension of the correlation hole 
is substantially larger than $a$ for small
volume fractions and large particle charges, it is then 
a good approximation to consider only the leading
Oseen-term in $\ve{B}(\ve{r})$ 
(cf. eqs. (\ref{series.A},\ref{series.B},\ref{H.pairlevel}))
in calculating the sedimentation velocity. 
$U/U_{0}$ is then well approximated by
\begin{equation} \label{U.EHS}
    \frac{U}{U_{0}} = 1 + \frac{3\phi}{a^{2}} \int_{0}^{\infty} dr \, r \,
    h(r) = 1 + \frac{3\phi}{a^{2}} \tilde{H}(s=0).
\end{equation}
Here, $h(r)=g(r)-1$ is the total correlation function and
\begin{equation}
    \tilde{H}(s) = \int_{0}^{\infty} dr \, r \, h(r) \, e^{-sr}  
\end{equation}
denotes the Laplace transform of $r h(r)$. Next, we approximate $h(r)$ by the
total correlation function $h_{EHS}(r;\phi_{EHS})$ of the EHS model, 
evaluated at
the effective volume fraction $\phi_{EHS}$. With this approximation 
for $h(r)$ used in
eq. (\ref{U.EHS}), we readily obtain the result $U/U_{0} = 1 - p \, \phi^{1/3}$
with a fractional exponent $1/3$, and the parameter $p$ given by
\begin{equation} \label{a.EHS}
    p =  3 \, \phi_{EHS}^{2/3} \, \tilde{H}_{EHS}(z=0),
\end{equation}
where $\tilde{H}_{EHS}(z)$ is the Laplace transform of 
$y \, h_{EHS}(y;\phi_{EHS})$ with $y = r/a_{EHS}$ and $z = s \, a_{EHS}$. 

We notice
that $\phi_{EHS}$ and hence $p$ are indeed independent of $\phi$ and $Z \,\,
(\ge 100)$, provided that $a_{EHS}$ is identified with $r_{m}/2$. This follows
from the fact that, for deionized suspensions of strongly charged particles
where $\kappa_{c}\gg\kappa_{s}$ holds,
$r_{m}$ coincides within $3\%$
with the average geometrical distance 
$\overline{r} = a {(3\phi/(4\pi))^{-1/3}}$ of two particles. 
For an illustration of this fact
consider the inset of fig. \ref{gr.rmsa.ehs.plot}, 
which shows RMSA results for $r_{m}/\overline{r}$ 
as a function of $\phi$. Thus, the scaling relation
$r_{m} \approx \overline{r} \propto \phi^{-1/3}$ holds 
and this gives rise to the
exponent $1/3$ in eq. (\ref{U.charged}): By identifying $a_{EHS}$ with
$\overline{r}$, we obtain an effective volume fraction
$\phi_{EHS} = \pi/6$ independent of $Z$ and $\phi$. 

To obtain a numerical value of $p$ according to eq. (\ref{a.EHS}), we
take now advantage of an analytic expression for 
$\tilde{H}_{EHS}(z)$ given in
the Percus-Yevick (PY) approximation \cite{Wertheim:63}. 
By performing the zero-$z$ limit,
we obtain in PY approximation after a straightforward calculation
the intermediate result
\begin{equation} \label{Htilde.PY}
    \tilde{H}_{EHS}(z=0) = -
    \frac{10-2 \, \phi_{EHS}+\phi_{EHS}^{2}}{5 \, (1+2 \, \phi_{EHS})}.
\end{equation}
Substitution of eq. (\ref{Htilde.PY}) into eq. (\ref{a.EHS}) gives
\begin{equation} \label{a.PY}
    p = \frac{3}{5} \, \phi_{EHS}^{2/3} \, 
        \frac{10-2 \, \phi_{EHS}+\phi_{EHS}^{2}}{1+2 \, \phi_{EHS}}.
\end{equation} 
Since $\phi_{EHS}=\pi/6$, we obtain from eq. (\ref{a.PY}) a 
value $p = 1.76$ remarkably close to 
$p=1.80$ as determined from the parametric fit of our 
numerical PA-result for $U/U_{0}$
(cf. following section). 

For an illustration of the replacement of the actual $g(r)$ by the
EHS-$g(r)$ within the 
EHS model, see again fig. \ref{gr.rmsa.ehs.plot},
which shows besides a typical RMSA-$g(r)$ of charge-stabilized
particles the corresponding 
EHS-$g(r)$ obtained in PY approximation for $\phi_{EHS}=\pi/6$.

We finish this section with two remarks. 
First, we would like to stress again the close connection between
the relation eq. (\ref{U.charged}), found for highly charged 
colloidal suspensions, and the corresponding result 
for ordered arrays of fixed spheres. For both systems,
the $\phi^{1/3}$-scaling behavior of $U/U_{0}$ is caused 
by a strong structural correlation of the particles, resulting in the
scaling relation $r_{m}\propto\phi^{-1/3}$ 
\cite{Naegele:94:1,Naegele:95:1,Naegele:Habil:published}.
Although Saffmann already mentioned the possibility for
finding a scaling behavior as eq. (\ref{U.charged}) for highly correlated
fluid suspensions \cite{Saffman:73}, the numerical method
descibed here provides the first quantitative results 
on the sedimentation velocity of charged suspensions, leading to the 
predicted $\phi^{1/3}$-scaling in eq. (\ref{U.charged}) 
plus the calculation of the prefactor $p$.
It is further important to notice 
that the scaling relation $r_m\propto\phi^{-1/3}$ is not valid if significantly
large amounts of excess electrolyte are added to the suspension,  
since then the particle diameter 
becomes another physically relevant length scale besides the mean particle
distance $\bar{r}$. The $\phi$-dependence of $U/U_0-1$ at small $\phi$ changes
then gradually with increasing $n_s$ from a $\phi^{1/3}$-dependence
to the linear $\phi$-dependence of eq. (\ref{U.hs}) 
characteristic for hard spheres \cite{ThiesWeesie:95}. 

\subsection{Weakly charged particles}

It is also of interest to consider the opposite limiting 
case of dilute suspensions
of weakly charged spheres in the framework of the effective macroion fluid
model. This case was addressed first by Petsev and Denkov  
\cite{Petsev:Denkov:92,Denkov:Petsev:92}.
Suppose $Z$ is so small that
\begin{equation} 
   \beta u_{el}(r=2a) = K \ll 1,
\end{equation}
i.e. the electrostatic repulsion can be treated as a small perturbation
of the hard-core repulsion between the particles. 
For very small $\phi$, $g(r)$ can then be approximated as
\begin{equation} \label{gr.linearized}
   g(r)\simeq g_0(r)\left[1-\beta u_{el}(r)\right]
\end{equation}
with $g_0(r)=\Theta(r-2a)$. In other words, $g(r)$ is approximated
by its zero-density limit 
$
   g(r)\simeq \exp\left(-\beta u(r)\right)
       = g_{0}(r)\exp\left(-\beta u_{el}(r)\right)
$, 
linearized with respect to $\beta u_{el}(r)$.
Substitution of eq. (\ref{gr.linearized}) in
eqs. (\ref{U.lim.H},\ref{H.pairlevel}) gives two additive contributions
to $U/U_0$, the first one arising from the hard-core part $g_0(r)$, and
the second one from the electrostatic perturbation of $g(r)$ in
eq. (\ref{gr.linearized}).
Explicitly \cite{Petsev:Denkov:92,Denkov:Petsev:92}
\begin{equation} \label{U.petsev}
   \frac{U}{U_0}\simeq 
      1-\phi\left[6.55+6\frac{K}{\kappa a}\right],
\end{equation}
where the first term in the bracket is the 
result of eq. (\ref{U.hs}), arising from the hard-core
part of $g(r)$. As far as the electrostatic part is concerned, it was
argued by Petsev and Denkov \cite{Petsev:Denkov:92}
that the leading Oseen-term in 
$\ve{B}(\ve{r})$ should give the dominant contribution to $U/U_0$.
Therefore only the Oseen-term is considered here, leading to the second term
in the bracket of eq. (\ref{U.petsev}).
We will provide a critical discussion of 
this approximation in the following section.

Next, we further simplify eq. (\ref{U.petsev}) for 
the case of vanishing excess 
electrolyte ($n_s=0$), where 
$\kappa a=A\phi^{1/2}$ with $A=(3|Z|L_B/a)^{1/2}$. When $\phi$ and/or
$|Z|$ are sufficiently small so that $A\phi^{1/2}\ll1$, eq.
(\ref{U.petsev}) is further simplified to 
\begin{equation} \label{U.phi.sqrt}
   \frac{U}{U_0}\simeq 
      1-\left[6\left(\frac{L_B}{2a}\right)|Z|^3\right]^{\frac{1}{2}}
        \phi^{\frac{1}{2}}+\cal{O}(\phi).
\end{equation}
Thus, we expect the reduced sedimentation velocity of dilute deionized
suspensions of weakly charged particles to scale like $U/U_0=1-p\phi^{1/2}$,
with a parameter $p$ depending on the macroion charge $Z$ 
and on the ratio $L_B/(2a)$.

Notice that the hard core contribution $6.55\phi$ to 
$U/U_0$ in eq. (\ref{U.petsev})
is omitted against the term proportional $\phi^{1/2}$ in proceeding from 
eq. (\ref{U.petsev}) to eq. (\ref{U.phi.sqrt}).
Therefore, eqs. (\ref{U.petsev}) and (\ref{U.phi.sqrt}) do not 
become equal to each other for $Z\to 0$. Whereas
eq. (\ref{U.petsev}) reduces to the hard sphere result of eq. (\ref{U.hs}) for
$Z=0$, eq. (\ref{U.phi.sqrt}) results in $U=U_0$ for vanishing particle charge.
Obviously, the range of
validity of eq. (\ref{U.phi.sqrt}) is restricted to small values of $Z$, 
but still sufficiently large that the second term in eq. 
(\ref{U.petsev}) plays the dominant
role compared to the hard core contribution $6.55\phi$.
We mention already here that due to the strong approximations made in deriving
eqs. (\ref{U.petsev},\ref{U.phi.sqrt}),
the range of validity of these two equations is restricted to 
extremely small values of $\phi$ and $Z$ (cf. following section).  

For eqs. (\ref{U.petsev},\ref{U.phi.sqrt}) to be valid, 
it is further assumed that the
van der Waals attraction among the particles is negligibly small.
Experimentally, this might be achieved by coating the 
particle surface with a thin layer of polymer chains and/or by matching the
solvent refractive index to the refractive index of the colloidal particles.
If attractions between the particles
are present, not only eqs. (\ref{U.petsev},\ref{U.phi.sqrt}) cease
to be valid, but also sedimentation velocities larger than
for hard spheres may be measured at the same density
\cite{Cheng:55,Buscall:82,Jansen:86:4,ThiesWeesie:95,ThiesWeesie:96}. 
This enhanced sedimentation rate is due
to the enhanced probability, as compared to hard spheres, 
of two particles for being close together, 
thus leading to a reduced retardation from
backflow \cite{Russel:book}.  

\section{Numerical results and discussion}
\label{results.section}

In this section, we present numerical results for the reduced 
sedimentation velocity $U/U_0$ as a function of the volume fraction $\phi$ 
and of the particle charge number $Z$. Our PA-scheme 
calculations of $U/U_{0}$ account, if not stated differently, for two-body
contributions to the hydrodynamic mobility tensors up to order
$(a/r)^{20}$. The system parameters employed in our calculations
are $\epsilon=2.183$ (corresponding to cis-decalin as dispersing
fluid at temperature $T=293$K), and particle radius $a=695$nm,
representing modified PMMA particles investigated very recently in
sedimentation experiments \cite{Ackerson:private}. 
We further use $n_s=0$, i.e. the ionic strength is
essentially determined by the (monovalent) counterions, which
counterbalance the charge of the colloidal particles.
Obviously, the dielectric constant $\epsilon$ used in the calculations presented
here is rather small, leading to strong long-ranged repulsions between
the particles even for small surface charges, which, most probably, are 
present in any polar organic solvent \cite{Philipse:review:97}.
Furthermore, for these solvents, residual water can cause strong
electric charging of the PMMA particles \cite{Ackerson:private}. 
We therefore present calculations for both
weakly and strongly charged particles.
Our calculations for strongly charged particles recover
the results obtained in Refs. 
\cite{Naegele:94:1,Naegele:95:1,Naegele:Habil:published} for
different system parameters, demonstrating clearly the independence of
the scaling relation eq. (\ref{U.charged}) from specific
system parameters, in particular from the solvent dielectric constant $\epsilon$.

\subsection{Strongly charged particles}

In fig. \ref{U.highcharge.plot}, we
show the PA-scheme result for $U/U_0$, obtained by
choosing an effective charge number $Z=150$ large enough that the 
physical hard-core radius $a$ of the particles constitutes no
relevant physical length scale. This numerical result is
perfectly fitted by the form $1-p\phi^\alpha$, with
$p=1.80$ and $\alpha=0.34\simeq1/3$.
As mentioned before, this is in remarkably good 
{\em quantitative} agreement with our EHS model result obtained
in PY approximation.
As discussed in detail in Ref. \cite{ThiesWeesie:95}, adding small
amounts of excess electrolyte leads to a significant increase in $U$,
and the $\phi^{1/3}$-scaling behaviour of $U/U_0$ does not hold any more.
For comparison, fig. \ref{U.highcharge.plot} includes also the reduced
sedimentation velocity of
uncharged hard spheres according to eq. (\ref{U.hs}). 
Evidently, the sedimentation
velocity of charged particles decreases much faster with 
increasing $\phi$ than one would
expect for hard spheres at the same volume fraction.
%
%
%
%

As shown in fig. \ref{gr.rmsa.ehs.plot}, the RMSA-$g(r)$ corresponding
to the largest volume fraction
$\phi=0.08$ considered in fig. \ref{U.highcharge.plot} 
has well developed undulations, with the
maximum value approximately located at the mean geometrical 
particle distance $\overline{r}$.
We only quote that the corresponding static structure factor $S(q)$ has its 
principal peak height well below the range of values 
$2.8-3.1$, where the system starts to
freeze according to the empirical Hansen-Verlet rule
\cite{Hansen:Verlet:69}.
A few comments on the quality of the RMSA input
for $g(r)$, used in the PA-scheme calculations, is in order here.
It is well known \cite{Naegele:Habil:published} that
the RMSA underestimates to some extend the oscillations of $g(r)$ and $S(q)$
in case of strongly correlated systems of highly charged particles. 
In principle,
we could use instead of the RMSA an alternative scheme like
the Rogers-Young (RY) scheme \cite{Rogers:Young:84}. The RY sheme is
quite accurate within the effective macrofluid model, but has the
disadvantage of being numerically far more involved than the RMSA.
Fortunately, the RMSA has been found to give nearly identical results
as the RY-scheme, provided that a
somewhat larger value of $Z$ is used in the RMSA calculations 
\cite{Naegele:Habil:published}.
We have argued before and will show in the following that 
the parameter $p$ in eq.
(\ref{U.charged}) is nearly independent of $Z$, typically as long as $Z\ge100$.
Consequently, using the RY-scheme instead of the RMSA for the
same value of $Z$ should lead to practically identical results for
$U/U_0$. We have verified this assertion by explicit RY calculations
\cite{Watzlawek:diploma, Watzlawek:Loehle:Naegele:unpublished}. 

In fig. \ref{U.Z.plot}, we display PA-scheme results for 
$U/U_0$ versus $\phi$ for various values of $Z$. For $Z=0$, we recover the
hard sphere result $U/U_0=1-p\phi$ with $p=6.54$.
An increase in $Z$ leads to a strong reduction in $U$, 
with a gradual transition
from the linear $\phi$-dependence of $U/U_0$ towards the non-linear form of 
eq. (\ref{U.charged}), with $p\simeq1.80$. This figure nicely illustrates that
$U/U_0$ becomes independent of $Z$ for $Z\ge100$.
For these large values of $Z$, $r_m$ stays practically constant when
$Z$ is increased with fixed $\phi$ \cite{Watzlawek:diploma}.
%
%
%
%

The effect on the PA-results for $U/U_0$ caused by truncating
the $(a/r)$-expansion of the two-particle mobilities after
various terms of increasing order in $(a/r)$, can be assessed from
fig. \ref{U.divcontrib.plot}. The solid line represents the
full PA-scheme result where all two-body contributions up to
${\cal O}(r^{-20})$ are accounted for. Nearly identical results for
$U/U_{0}$ are obtained, even at
$\phi=0.08$, when contributions only up to ${\cal O}(r^{-4})$
are considered. In contrast, the result for $U/U_{0}$ obtained by 
accounting only for
hydrodynamic contributions up to ${\cal O}(r^{-3})$ shows clear
deviations from the full PA-scheme result at volume fractions
$\phi\ge 0.01$.
As a conclusion, we can state that for dispersions 
of highly charged particles it
is justified to use a truncated far-field
expansion of the mobility tensors with only the
first terms being included, provided a good static input for $g(r)$ is used.
The finding that an expansion up to ${\cal O}(r^{-4})$ leads
already to good results
indicates further that hydrodynamic $n$-body contributions to $U$ with
$n\ge3$ are indeed negligibly small in the considered $\phi$-range.
For practical purposes, it is therefore suitable to use the simple eq.
(\ref{U.linear.phi}) for calculating the reduced sedimentation
velocity for highly charged suspensions.
%
%
%
%

\subsection{Weakly charged particles}

We discuss now the behaviour of $U/U_0$ when $Z$ is small. To be
specific, consider first the value $Z=5$. For this charge number,
$\beta u_{el}(r)<(L_B/(2a))Z^2\simeq0.5$. This
implies that the microstructure is 
affected not only by the electrostatic repulsion but also by the physical 
hard core of the particles. In fig. \ref{U.div.gr.plot}, we have redrawn
from fig. \ref{U.Z.plot} the PA-scheme result for $U/U_0$ with $Z=5$ and
RMSA-input for $g(r)$. 
%
%
%
%
This graph should be compared with the corresponding 
result for $U/U_0$ obtained from the expression given in
eq. (\ref{U.petsev}) (henceforth referred to as PD-result). 
The PD-result for $U/U_0$ is completely different from the corresponding 
PA-scheme result, in particular at larger $\phi$ where the PD-$U/U_0$
even turns negative. The failure of the PD-expression in describing
$U/U_{0}$ arises from the fact 
that the approximations entering in its derivation, i.e. in particular
the approximation eq. (\ref{gr.linearized}) for $g(r)$, but also the
truncation of the electrostatic contribution to $U/U_0$ after the leading
Oseen-term, are not justified even for a charge number as small as $Z=5$. 
The consideration of more terms in the 
hydrodynamic pair mobility contributions up to 
${\cal O}(r^{-20})$, however, leads only to a small increase in $U/U_0$, 
as can be seen from fig. \ref{U.div.gr.plot}, which further
includes the PA-scheme result 
for $U/U_{0}$ with $g(r)$ approximated by the linearized zero-density form of
$g(r)$, eq. (\ref{gr.linearized}). Another slight improvement in $U/U_0$
is achieved when the non-linearized zero-density form 
$g(r)=\Theta(r-2a)\exp\left(-\beta u_{el}(r)\right)$ is used as
input in the PA-scheme. In fact, the contact value $\beta u_{el}(2a)$
is not small enough for $Z=5$ to fully justify a linearization
in $\beta u_{el}(r)$. Therefore, the main reason for the failure of the
PD-expression is due to the significant differences of the actual $g(r)$ at
$Z=5$ from its zero-density form already
for $\phi\simeq0.01$.

To obtain further insight in the range of validity of the PD-result eq. 
(\ref{U.petsev}), we display in fig. \ref{U.Zsmall.plot} results for
$U/U_0$ for an even smaller particle charge $Z=2$ and 
volume fractions smaller than $\phi=0.01$. 
%
%
%
%
As seen from this figure and from fig.
\ref{U.div.gr.plot}, the 
differences between the RMSA-PA-scheme result for $U/U_{0}$ 
and the PA-scheme result for using 
$g(r)=\Theta(r-2a)\exp(-\beta u_{el}(r))$ or 
$g(r)=\Theta(r-2a)\left[1-\beta u_{el}(r)\right]$
as static input become smaller with decreasing $Z$,
but are still significant even 
for $Z=2$ and $\phi\stackrel{>}{\sim}10^{-3}$.
For an explanation of this finding, we refer to fig. \ref{gr.Z2.plot},
where the radial distribution $g(r)$ obtained from the RMSA and from the 
zero-density expression $g(r)=\Theta(r-2a)\exp(-\beta u_{el}(r))$ 
for $Z=2$ and 
$\phi=0.005$ are shown. 
%
%
%
%
Obviously, there are still small
differences between the the zero-density expression
for $g(r)$ and the corresponding RMSA result even for these small values
of $Z$ and $\phi$.
We show now that these small differences in $g(r)$ cause the large differences
in the results for $U/U_0$ as illustrated in fig. \ref{U.Zsmall.plot}.
For this purpose, we plot in the inset of fig. \ref{gr.Z2.plot}
the function $r(1-g(r))$ as obtained from the
radial distribution functions shown in the main figure.
As easily seen, the two curves for $r(1-g(r))$, corresponding to the two
different inputs for $g(r)$, are remarkably different
from each other even up to large distances $r$.
It is now crucial to notice that according to eq. (\ref{U.linear.phi})
it is essentially the function $r(1-g(r))$ and not $g(r)$ itself
which appears in the integrand when $U/U_0$ is calculated including
terms up to ${\cal O}(r^{-1})$
in the series expansions in eqs. (\ref{series.A},\ref{series.B}). 
Therefore, the small differences in the two $g(r)$'s considered
in fig. \ref{gr.Z2.plot} give rise to substancial differences in the
corresponding sedimentation velocities shown in fig. \ref{U.Zsmall.plot}.
We can therefore conclude that it
is absolutely important to employ an accurate $g(r)$-input for calculating
$U/U_0$ in the PA-scheme. Due to the use
of the (linearized) zero-density approximation of $g(r)$,
the expression given by Petsev and Denkov in eq. (\ref{U.petsev})
does not predict the sedimentation velocity of 
weakly charged particles correctly.     
  
After having explored the high sensibility of $U/U_0$ on the 
form of the $g(r)$-input used in the PA-scheme, it is
now apparent that the second approximation made in deriving
eq. (\ref{U.petsev}), i.e. the omission of terms in the far-field expansions
of the hydrodynamic mobilities 
of ${\cal O}(r^{-3})$ in the electrostatic
contribution to $U/U_0$, is not valid for
the small values of $Z$ and $\phi$ used in figs. \ref{U.Zsmall.plot} and
\ref{gr.Z2.plot}. This conclusion follows from fig. \ref{U.Zsmall.plot},
when the results for $U/U_{0}$ according to eq. (\ref{U.petsev}) 
and derived from the PA-scheme using eq. (\ref{gr.linearized}) as static input
are compared. Although the same approximation 
is employed for $g(r)$, the two results for $U/U_{0}$ do not agree,
since the approximations in truncating the series expansions of the
hydrodynamic mobility tensors after the terms of ${\cal O}(r^{-1})$ and
${\cal O}(r^{-20})$, respectively, are not the same.
On the other hand, we have shown in fig. \ref{U.divcontrib.plot} 
for highly charged particles
that the first few terms in the series expansion of
eqs. (\ref{series.A},\ref{series.B}) are sufficient to calculate
$U/U_0$. As pointed out before, this is due to the existence of 
an extended correlation hole in the $g(r)$ of highly charged 
particles in deionized
suspensions (cf. fig. \ref{gr.rmsa.ehs.plot}). The correlation hole
gives rise to a fast concergence of the integrals in 
eq. (\ref{H.pairlevel}).
However, in case of the weakly charged particles considered here, 
there is no correlation hole present, as can be seen, e.g., from fig. 
\ref{gr.Z2.plot}. Therefore, a large number of terms in the series expansions
of eqs. (\ref{series.A},\ref{series.B}) are needed for
weakly charged particles to calculate $U/U_{0}$. 
In fact, as discussed before in the case of hard spheres,
the terms of ${\cal O}(r^{-3})$ and ${\cal O}(r^{-4})$ in 
the series expansions
in eqs. (\ref{series.A},\ref{series.B}) contribute in approximately 
the same weight to $U/U_0$, still remarkably large compared the the 
leading Oseen-term of ${\cal O}(r^{-1})$, 
considered only in the electrostatic contribution in the 
PD-result eq. (\ref{U.petsev}).

To summarize the last presented results, we have shown that
the range of validity of eq. (\ref{U.petsev}) is restricted to
values of $\phi$ considerably smaller than $10^{-3}$, where the
macroion radial distribution function is described very accurately
by its linearized zero-density form.   
Furthermore, for the PD-result eq. (\ref{U.petsev}) to be hold, 
the effective charge $Z$ has to be small enough that the
approximation in eq. (\ref{gr.linearized})
can be used to calculate $U/U_0$, but still not so small
that the actual $g(r)$ is
too close to its zero-charge limiting form $g_0(r)=\Theta(r-2a)$. 
Otherwise, the omission
of higher order hydrodynamic terms of ${\cal O}(r^{-3})$ to the
electrostatic contribution in eq. (\ref{U.petsev})
ceases to be a good approximation.
These severe restrictions limit the range of
validity of eq. (\ref{U.petsev}) to values of the system parameters
most probably not accessible in sedimentation experiments.
Moreover, the relative differences in $U$ and $U_{0}$ become
small in the parameter range where eq. (\ref{U.petsev}) should apply. 

Let us now turn to some remarks on the 
$\phi^{1/2}$-scaling of $U/U_0-1$ proposed in eq. (\ref{U.phi.sqrt}).
Since eq. (\ref{U.phi.sqrt}) is 
derived from the PD-result eq. (\ref{U.petsev}),
the before discussed restricted range of validity of eq. (\ref{U.petsev})
applies also to eq. (\ref{U.phi.sqrt}).
Furthermore, in eq. (\ref{U.phi.sqrt}), the terms linear
in $\phi$ are neglected against the term proportional to $\phi^{1/2}$.
We have shown in fig. \ref{U.Zsmall.plot} that this additional approximation
becomes invalid already at small volume fractions, typically
$\phi\stackrel{>}{\sim}10^{-3}$. For larger $\phi$, 
the two results for $U/U_{0}$ in eq. (\ref{U.petsev}) 
and eq. (\ref{U.phi.sqrt}) start
 to deviate strongly, since the 
contributions to $U/U_0$ linear in $\phi$ (especially the hard core 
contribution $6.55\phi$) are no longer small 
compared to the term proportional to $\phi^{1/2}$.
We therefore also expect the $\phi^{1/2}$-scaling of $U/U_0-1$ according
to eq. (\ref{U.phi.sqrt}) not to be measurable in a sedimentation experiment.
Nevertheless, we wish to point out that there
is a specific range of particle charges $Z$ where $U/U_0-1$ 
indeed scales as $\phi^{1/2}$
for a broad range of volume fractions. This range of $Z$-values
is determined when the effective 
particle charge is subsequently lowered from large values $Z\ge 100$, 
where  $U/U_0-1$ scales like $\phi^{1/3}$, to $Z=0$, where $U/U_0-1$ 
behaves linearly in $\phi$ according
to eq. (\ref{U.hs}). To demonstate the occurance of a $\phi^{1/2}$-dependence
of $U/U_0-1$, we have redrawn in 
fig. \ref{U.phisqrt.plot} the RMSA-PA result from fig.
\ref{U.div.gr.plot} for $Z=5$ together with the corresponding 
results for $Z=6$ and $Z=7$. 
%
%
%
%
Obviously, all three curves are well described by the form 
$U/U_0=1-p\phi^\alpha$ with a parameter
$\alpha$ very close to $1/2$.
Such a $\phi^{1/2}$-dependence of $U/U_0$ might indeed be 
measurable in sedimentation experiments on dispersions with 
suitably chosen particle charges. However, we  
stress that the occurance of the parameter $\alpha$ close to 
$1/2$ for a broad range of volume fractions as dipicted
in fig. \ref{U.phisqrt.plot}
is completely different from the prediction eq. (\ref{U.phi.sqrt}), which 
has been shown to be valid only for very small $\phi$.    

Finally, we shortly comment on the dependence of our results
on the system parameters $T$ and $\epsilon$, which are held fixed in 
our calculations.
Obviously, these two parameters enter into our calculations mainly by
determining the Bjerrum length $L_B$, which has a rather large value
$L_B=26.12$nm due to the small dielectric constant 
$\epsilon=2.183$ used in this work.
As consequence, also the ratio $L_B/(2a)$ is comparatively large.
This ratio determines together with $Z$ the strength
of the electrostatic repulsion between the particles.
For that reason, one might argue that our dicussion concerning the
use of eq. (\ref{gr.linearized}) for $g(r)$ in calculating 
$U/U_0$ is not longer valid for smaller values of $L_B$. 
However, explicit PA-scheme calculations for system parameters as
used in Ref. \cite{ThiesWeesie:95} and/or a simple estimate show that
our conclusions derived above remain valid even for
considerably smaller values of $L_B$:
Consider a value of $L_{B}$ ten times smaller than the one used in the
above presented
calculations, e.g. a value like in 
sedimentation experiments on charged silica particles in ethanol 
\cite{ThiesWeesie:95}.
Then, according to eqs. (\ref{yukawa.potential},\ref{yukawa.coupling}),
one might choose a particle charge approximately 
$\sqrt{10}\simeq 3.16$ times larger than in our
calculations to achieve the same strength of repulsion between
the particles. As shown above for our choice of system parameters, 
the use of eq. (\ref{gr.linearized}) for $g(r)$
is a poor approximation even for low volume fractions $\phi\le 0.01$
and a particle charge $Z=2$. Therefore, even for
a system with $L_{B}$ ten times larger, eqs. (\ref{U.petsev},\ref{U.phi.sqrt})
are applicable only for particle charges considerably smaller than $Z=6$.
This slightly extended parameter range for eqs. 
(\ref{U.petsev},\ref{U.phi.sqrt}) to be valid is most probably still too
restricted to be experimentally accessible.

\section{Concluding remarks}

We have presented theoretical results for the reduced sedimentation velocity 
$U/U_0$ of monodisperse charged suspensions in dependence of the volume fraction 
$\phi$ and of the particle charge number $Z$. Our theoretical
model for $U/U_0$ is based on the effective macroion fluid model and on the
assumption that pairwise additive HI prevails at sufficiently small
$\phi$. The numerical results for $U/U_0$ in case of 
strongly charged particles at
low salinity are well parametrized by the form $1-p\phi^{1/3}$.

We have shown that the exponent $1/3$ and the value of the charge-independent
parameter $p = 1.8$ can be {\em quantitatively}
understood in terms of a 
model of effective hard spheres
of radius $a_{EHS}$ which depends on the volume fraction. Using 
Percus-Yevick input for the static pair correlation function of hard spheres,
the EHS model can be treated analytically and leads to a value of $p$ very close to $1.8$.

It was further demonstrated that $U/U_0-1$ can scale like
$\phi^{1/2}$ in case of dilute suspensions of very weakly charged particles.
This peculiar volume fraction dependence of $U/U_0$ 
derives from an expression given
by Petsev and Denkov \cite{Petsev:Denkov:92,Denkov:Petsev:92},
when it is further assumed that the ionic strength in the system is
mainly due to counterions. Our numerical calculations reveal
that both the original expression for $U/U_0$ given by Petsev and Denkov, and
the derived expression showing the $\phi^{1/2}$-scaling of $U/U_0-1$,
are only valid for very small particle volume fractions, which we expect not
to be accessible in conventional sedimentation experiments. 

\section*{Acknowledgements}

We are grateful to Barbara L{\"o}hle (University of Konstanz) for 
providing Rogers-Young calculations of various static distribution
functions, and to Bruce Ackerson (Oklahoma State University) and Barbara Mandl
(formerly University of Konstanz) for helpful discussions.
We further thank on of the referees for calling our attention to
Ref. \cite{Saffman:73}.
M.W. thanks the Deutsche Forschungsgemeinschaft for financial support
within SFB 513 and SFB 237.

\newpage

%

\begin{thebibliography}{10}

\bibitem{Batchelor:72}
{G.~K. Batchelor},
\newblock {\em J. Fluid Mech.} {\bf 52}, 245 (1972).

\bibitem{Cichocki:89}
{B.~Cichocki} and {B.~U. Felderhof},
\newblock {\em Physica} {\bf A 154}, 213 (1989).

\bibitem{Clerx:92}
{H.~J.~H. Clerx} and {P.~P. J.~M. Schram},
\newblock {\em J. Chem. Phys.} {\bf 96}, 3137 (1992).

\bibitem{Kops-Werkhoven:81}
{M.~M. Kops-Werkhoven} and {H.~M. Fijnaut},
\newblock {\em J. Chem. Phys.} {\bf 74}, 1618 (1981).

\bibitem{Buscall:82}
{R.~Buscall}, {J.~W. Goodwin}, {R.~H. Ottewill}, and {T.~F. Tadros},
\newblock {\em J. Coll. Int. Sci.} {\bf 85}, 78 (1982).

\bibitem{ThiesWeesie:96}
{D.~M.~E. Thies-Weesie}, {A.~P. Philipse}, and {H.~N.~W. Lekkerkerker},
\newblock {\em J. Coll. Int. Sci.} {\bf 177}, 427 (1996).

\bibitem{Xu:98}
{W.~Xu}, {A.~Nikolov}, and {D.~T. Wasan},
\newblock {\em J. Coll. Int. Sci.} {\bf 197}, 160 (1998).

\bibitem{Reed:Anderson:76}
{C.~C. Reed} and {J.~L. Anderson},
\newblock in {\em Hydrosols and Rheology}, edited by {M.~Kerker}, volume~4 of
  {\em Colloid and Interface Science}, Academic Press, New York,
  1976.

\bibitem{Reed:Anderson:80}
{C.~C. Reed} and {J.~L. Anderson},
\newblock {\em AIChE J.} {\bf 26}, 816 (1980).

\bibitem{Snook:84}
{I.~Snook} and {W.~van Megen},
\newblock {\em J. Coll. Int. Sci.} {\bf 100}, 194 (1984).

\bibitem{Dickinson:80}
{E.~Dickinson},
\newblock {\em J. Coll. Int. Sci.} {\bf 73}, 578 (1980).

\bibitem{Pusey:Tough:85}
{P.~N. Pusey} and {R.~J.~A. Tough},
\newblock in {\em Dynamic light scattering}, edited by {R.~Pecora}, 
  Plenum, New York, 1985.

\bibitem{Pusey:91}
{P.~N. Pusey},
\newblock in {\em Liquids, Freezing and Glass Transition: II}, edited by {J.-P.
  Hansen}, {D.~Levesque}, and {J.~Zinn-Justin}, North Holland, Amsterdam, 1991.

\bibitem{Naegele:94:1}
{G.~N{\"a}gele}, {B.~Steininger}, {U.~Genz}, and {R.~Klein},
\newblock {\em Physica Scripta} {\bf T 55}, 119 (1994).

\bibitem{Naegele:Habil:published}
{G.~N{\"a}gele},
\newblock {\em Phys. Rep.} {\bf 272}, 215 (1996).

\bibitem{ThiesWeesie:95}
{D.~M.~E. Thies-Weesie}, {A.~P. Philipse}, {G.~N{\"a}gele}, {B.~Mandl}, and
  {R.~Klein},
\newblock {\em J. Coll. Int. Sci.} {\bf 176}, 43 (1995).

\bibitem{Naegele:93}
{G.~N{\"a}gele}, {O.~Kellerbauer}, {R.~Krause}, and {R.~Klein},
\newblock {\em Phys. Rev. E} {\bf 47}, 2562 (1993).

\bibitem{Naegele:95:1}
{G.~N{\"a}gele}, {B.~Mandl}, and {R.~Klein},
\newblock {\em Progr. Colloid Polym. Sci.} {\bf 98}, 117 (1995).

\bibitem{Baur:96:1}
{P.~Baur}, {G.~N{\"a}gele}, and {R.~Klein},
\newblock {\em Phys. Rev. E} {\bf 53}, 6224 (1996).

\bibitem{Naegele:97:1}
{G.~N{\"a}gele} and {P.~Baur},
\newblock {\em Physica} {\bf A 245}, 297 (1997).

\bibitem{Ackerson:private}
{B.~J. Ackerson},
\newblock Oklahoma State University,
\newblock private communication.

\bibitem{Watzlawek:96:1}
{M.~Watzlawek} and {G.~N{\"a}gele},
\newblock {\em Physica} {\bf A 235}, 56 (1997).

\bibitem{Watzlawek:97:1}
{M.~Watzlawek} and {G.~N{\"a}gele},
\newblock {\em Prog. Coll. Polym. Sci.} {\bf 104}, 168 (1997).

\bibitem{Watzlawek:97:2}
{M.~Watzlawek} and {G.~N{\"a}gele},
\newblock {\em Phys. Rev. E} {\bf 56}, 1258 (1997).

\bibitem{Hasimoto:59}
{H.~Hasimoto},
\newblock {\em J. Fluid Mech.} {\bf 5}, 317 (1959).

\bibitem{Saffman:73}
{P.~G. Saffman},
\newblock {\em Stud. Appl. Math.} {\bf 52}, 115 (1973).

\bibitem{Zick:Homsy:82}
{A.~A. Zick} and {G.~M. Homsy},
\newblock {\em J. Fluid Mech.} {\bf 115}, 13 (1982).

\bibitem{Brady:88}
{J.~F. Brady}, {R.~J. Phillips}, {J.~C. Lester}, and {G.~Bossis},
\newblock {\em J. Fluid Mech.} {\bf 195}, 257 (1988).

\bibitem{Wertheim:63}
{M.~S. Wertheim},
\newblock {\em Phys. Rev. Lett.} {\bf 10}, 321 (1963).

\bibitem{Jones:Pusey:91}
{R.~B. Jones} and {P.~N. Pusey},
\newblock {\em Annu. Rev. Phys. Chem.} {\bf 42}, 137 (1991).

\bibitem{Dhont:book}
{J.~Dhont},
\newblock {\em An Introduction to Dynamics of Colloids}, volume~2 of {\em
  Studies in interface science (eds. D. M{\"o}bius and R. Miller)},
\newblock Elsevier, Amsterdam, 1996.

\bibitem{Russel:81}
{W.~B. Russel} and {A.~B. Glendinning},
\newblock {\em J. Chem. Phys.} {\bf 74}, 948 (1981).

\bibitem{Russel:book}
{W.~B. Russel}, {D.~A. Saville}, and {W.~R. Schowalter},
\newblock {\em Colloidal Dispersions},
\newblock Cambridge University Press, Cambridge, 1989.

\bibitem{Glendinning:82}
{A.~B. Glendinning} and {W.~B. Russel},
\newblock {\em J. Coll. Int. Sci.} {\bf 89}, 124 (1982).

\bibitem{Ladd:90}
{A.~J.~C. Ladd},
\newblock {\em J. Chem. Phys.} {\bf 93}, 3484 (1990).

\bibitem{Ladd:93}
{A.~J.~C. Ladd},
\newblock {\em Phys. Fluids A} {\bf 5}, 299 (1993).

\bibitem{Beenakker:Mazur:84}
{C.~W.~J. Beenakker} and {P.~Mazur},
\newblock {\em Physica} {\bf A 126}, 349 (1984).

\bibitem{Genz:91}
{U.~Genz} and {R.~Klein},
\newblock {\em Physica} {\bf 171}, 26 (1991).

\bibitem{Jones:Schmitz:88}
{R.~B. Jones} and {R.~Schmitz},
\newblock {\em Physica} {\bf A 149}, 373 (1988).

\bibitem{Cichocki:Felderhof:Schmitz:88}
{B.~Cichocki}, {B.~U. Felderhof}, and {R.~Schmitz},
\newblock {\em Physico\-Chem. Hydr.} {\bf 10}, 383 (1988).

\bibitem{Petsev:Denkov:92}
{D.~N. Petsev} and {N.~D. Denkov},
\newblock {\em J. Coll. Int. Sci.} {\bf 149}, 329 (1992).

\bibitem{Denkov:Petsev:92}
{N.~D. Denkov} and {D.~N. Petsev},
\newblock {\em Physica} {\bf A 183}, 462 (1992).

\bibitem{Cheng:55}
{P.~Y. Cheng} and {H.~K. Schachman},
\newblock {\em J. Polym. Sci.} {\bf 16}, 19 (1955).

\bibitem{Jansen:86:4}
{J.~W. Jansen}, {C.~G. {De Kruif}}, and {A.~Vrij},
\newblock {\em J. Coll. Int. Sci} {\bf 114}, 501 (1986).

\bibitem{Philipse:review:97}
{A.~P. Philipse},
\newblock {\em Curr. Opinion Coll. Int. Sci.} {\bf 2}, 200 (1997).

\bibitem{Hansen:Verlet:69}
{J.-P. Hansen} and {L.~Verlet},
\newblock {\em Phys. Rev.} {\bf 184}, 150 (1969).

\bibitem{Rogers:Young:84}
{F.~J. Rogers} and {D.~A. Young},
\newblock {\em Phys. Rev. A} {\bf 30}, 999 (1984).

\bibitem{Watzlawek:diploma}
{M.~Watzlawek},
\newblock Diplom--thesis,
\newblock University of Konstanz, 1996.

\bibitem{Watzlawek:Loehle:Naegele:unpublished}
{M.~Watzlawek}, {B.~L{\"o}hle}, and {G.~N{\"a}gele},
\newblock unpublished results.

\end{thebibliography}
%

%
%

\newpage

\section*{Figure captions}

\begin{figure}[hbt]
   \caption{ \label{gr.rmsa.ehs.plot} 
            RMSA radial distribution function $g(r)$ (dashed line) 
            for a suspension of charged particles with 
            system parameters $a=695$ nm, $\epsilon=2.183$,
            $T= 293$ K, $n_s=0$ (i.e. no added salt), $Z=150$, and $\phi=0.08$.
            Also shown is the radial distribution function of the
            EHS model in PY approximation (solid line). 
            The effective volume fraction
            is $\phi_{EHS}=\pi/6$, corresponding to the identification
            $2a_{EHS}=r_m=\overline{r}$ (see text). 
            The inset shows the peak position $r_{m}$ of the 
            RMSA-$g(r)$ for the same system paramters, normalized by the mean
            particle distance $\overline{r}$, as function of $\phi$. 
           }
\end{figure} 
\begin{figure}[hbt]
   \caption{ \label{U.highcharge.plot} 
            PA-scheme result for the reduced sedimentation velocity $U/U_0$    
            of strongly charged particles in a deionized solvent (solid line). 
            The result is well parametrized
            by the form $1 - 1.80  \, \phi^{0.34}$. 
            System parameters as in fig. \ref{gr.rmsa.ehs.plot}. 
            For comparison, the dashed-dotted line represents the 
            lowest-order density result for hard spheres according to eq.
            (\ref{U.hs}).
           }
\end{figure}
\begin{figure}[hbt]
   \caption{ \label{U.Z.plot} 
            PA-scheme results for $U/U_0$ for various values of the effective
            charge number $Z$ as indicated in the figure. All other system
            parameters as in fig. \ref{gr.rmsa.ehs.plot}. 
            Notice
            that $U/U_0$ is nearly independent of $Z$ when $Z\ge100$. 
           }
\end{figure}
\begin{figure}[hbt]
   \caption{ \label{U.divcontrib.plot} 
            Dependence of $U/U_0$ on various
            two-body contributions to the hydrodynamic mobilities.
            Included terms of the series expansions in eqs.
            (\ref{series.A},\ref{series.B}) and used in the
            PA-scheme as indicated in the 
            figure. System parameters as in fig. \ref{U.highcharge.plot}.
           }
\end{figure}

\newpage

\begin{figure}[hbt]
   \caption{ \label{U.div.gr.plot} 
            $U/U_{0}$ versus $\phi$ for particle charge $Z=5$. 
            Other parameters chosen as in fig. \ref{gr.rmsa.ehs.plot}. 
            Shown are the PA-scheme results obtained
            using as static input the RMSA-$g(r)$ (solid line), 
            the zero-density form
            $g(r)=\Theta(r-2a)\exp\left(-\beta u_{el}(r)\right)$
            (dashed line), and the linearized zero-density form of
            $g(r)$ according to (\ref{gr.linearized}) 
            (dashed-dotted line). Further displayed
            is $U/U_0$ corresponding to eq. (\ref{U.petsev}) (dotted line). 
           }
\end{figure}
\begin{figure}[hbt]
   \caption{ \label{U.Zsmall.plot} 
            $U/U_{0}$ versus $\phi$ for particle charge $Z=2$.
            All other parameters as in fig. \ref{gr.rmsa.ehs.plot}. 
            Shown are PA-scheme results obtained using as input the 
            RMSA-$g(r)$ (solid line), and the zero-density form
            $g(r)=\Theta(r-2a)\exp\left(-\beta u_{el}(r)\right)$
            (dashed line). The latter is practically identical with the
            PA-$U/U_{0}$ calculated with the linearized 
            $g(r)$ of eq. (\ref{gr.linearized}). 
            Further displayed
            are results for $U/U_0$ according to
            eq. (\ref{U.petsev}) (dashed-dotted line)
            and eq. (\ref{U.phi.sqrt}) (dotted line).
            The thin solid line represents the hard sphere result 
            of eq. (\ref{U.hs}). 
           }
'\end{figure}
\begin{figure}[hbt]
   \caption{ \label{gr.Z2.plot} 
            Radial distribution functions $g(r)$ for $Z=2$ and $\phi=0.005$, 
            obtained
            in the RMSA (solid line) and given by $g(r)=\exp(-\beta u(r))$
            (dashed line).
            Parameters aside from $Z$ and $\phi$ as in  
            fig. \ref{gr.rmsa.ehs.plot}.
            The inset shows the corresponding functions $r(1-g(r))$.
           }
\end{figure}
\begin{figure}[hbt]
   \caption{ \label{U.phisqrt.plot} 
            RMSA-PA-scheme results for $U/U_{0}$ at particle charges
            $Z=5, Z=6,$ and $Z=7$.
            All other parameters as in fig. \ref{gr.rmsa.ehs.plot}.
            The result for $Z=5$ (solid line) is well parametrized by the
            form $U/U_0=1-2.18\phi^{0.56}$, whereas 
            the results for $Z=6$ (dashed line)
            and $Z=7$ (dashed-dotted line) are very well fitted 
            within the shown volume fraction range by $U/U_0=1-2.06\phi^{0.53}$
            and $U/U_0=1-1.97\phi^{0.50}$, respectively.
           }
\end{figure}

\end{document}